\documentclass[sigconf]{acmart}
\AtBeginDocument{%
  }

\setcopyright{acmlicensed}
\copyrightyear{2018}
\acmYear{2018}
\acmDOI{XXXXXXX.XXXXXXX}

\acmConference[SIGIR'24]{}{July 14-18, 2024}{Washington D.C., USA}
\acmISBN{978-1-4503-XXXX-X/18/06}

\begin{document}

\title{Learn by Selling: Equipping Large Language Models with Product Knowledge for Context-Driven Recommendations}
\author{Sarthak Anand}
\email{sarthak.anand@ingka.ikea.com}
\affiliation{%
  \institution{Ingka Group}
  \country{Netherlands}
}

\author{Yutong Jiang}
\affiliation{%
  \institution{Ingka Group}
  \country{Sweden}}
\email{yutong.jiang2@ingka.ikea.com}

\author{Giorgi Kokaia}
\affiliation{%
  \institution{Ingka Group}
 \country{Netherlands}}
\email{giorgi.kokaia@ingka.ikea.com}

\renewcommand{\shortauthors}{Anand et al.}

\begin{abstract}
The rapid evolution of large language models (LLMs) has opened up new possibilities for applications such as context-driven product recommendations. However, the effectiveness of these models in this context is heavily reliant on their comprehensive understanding of the product inventory. This paper presents a novel approach to equipping LLMs with product knowledge by training them to respond contextually to synthetic search queries that include product IDs. We delve into an extensive analysis of this method, evaluating its effectiveness, outlining its benefits, and highlighting its constraints. The paper also discusses the potential improvements and future directions for this approach, providing a comprehensive understanding of the role of LLMs in product recommendations.
\end{abstract}





\maketitle

\section{Introduction}
The rapid advancement of large language models (LLMs) has revolutionized the field of natural language processing, enabling these models to comprehend and generate human-like language with unprecedented accuracy. As a result, LLMs have been increasingly explored for their potential applications in various domains, including e-commerce. One such application is contextual product recommendation, where the model is tasked with suggesting relevant products to users based on their search queries, browsing history, and preferences.

Traditional product recommendation systems rely on collaborative filtering, content-based filtering, or hybrid approaches, which often suffer from limitations such as cold start problems, sparsity, and lack of personalization. In contrast, LLMs, with their ability to understand nuances of language and context, offer a promising solution to overcome these limitations. However, a crucial prerequisite for LLMs to excel in product recommendation is their possession of comprehensive knowledge about the entire inventory of products available for sale.

In this paper, we propose a novel approach to equip LLMs with product knowledge by training them to generate contextual responses to synthetic search queries containing product IDs. This approach enables the model to learn the relationships between products, their attributes, and user preferences, allowing it to provide personalized and relevant product recommendations. We evaluate the efficacy of this approach, discussing its advantages and limitations, and explore the potential of LLMs in transforming the landscape of product recommendation systems.
\section{Related Work}
\subsection{Large Language Models}
Recently, substantial progress has been made in the development of large language models(LLMs). Among commercial endeavors, OpenAI's ChatGPT\cite{brown2020language, ouyang2022training, openai2024gpt4}, has emerged as a prominent example, demonstrating promising results across various applications. These kinds of models have gained widespread adoption, serving as a powerful tool for natural language understanding as a LLM-based chatbot. Concurrently, the open-source community has made significant contributions with models such as LLaMA\cite{touvron2023llama} and Mistral\cite{jiang2023mistral, jiang2024mixtral}. These models not only perform competitively on numerous benchmarks but also contribute to explainability of LLMs. The insights derived from such models are pivotal in advancing the field, particularly in aspects of explainability and transparency in models' behavior.

\subsection{LLM-based Recommendation}
The development and application of large language models(LLMs) have significantly impacted the field of artificial intelligence, prompting substantial interest in integrating these models with recommender systems. Predominantly, Predominantly, LLM-based recommender systems have been deployed for tasks such as rating or ranking prediction\cite{kojima2023large, kang2023llms, yue2023llamarec, Bao_2023} and sequential recommendation \cite{wang2023zeroshot}. More specifically, in many cases, with prompt engineering, language model takes user profile, historical information and retrieved candidates as input and output the predicted ranking\cite{yue2023llamarec}.

Furthermore, there is an ongoing discourse\cite{wang2024recmind} addressing the challenges of generalizability in LLM-based systems as current research often focuses on training and fine-tuning LLMs on task-specific datasets. These approaches like fine tuning, while effective in specialized contexts, sometimes limits the broader applicability of the models across diverse scenarios.

In our research, we have enhanced the traditional approach of merely utilizing product information for fine-tuning by incorporating specific explanations on product recommendations. This method leverages the advanced reasoning capabilities of large language models to elucidate the rationale behind each recommendation. During the fine-tuning phase, we experimented with the introduction of additional tokens to evaluate their impact on the model's performance. The results demonstrate that our model not only retrieves products efficiently but also provides clear, explicit reasoning for each recommendation.

\section{Dataset}
\begin{figure*}[]
    \centering
    \includegraphics[width=0.9\textwidth]{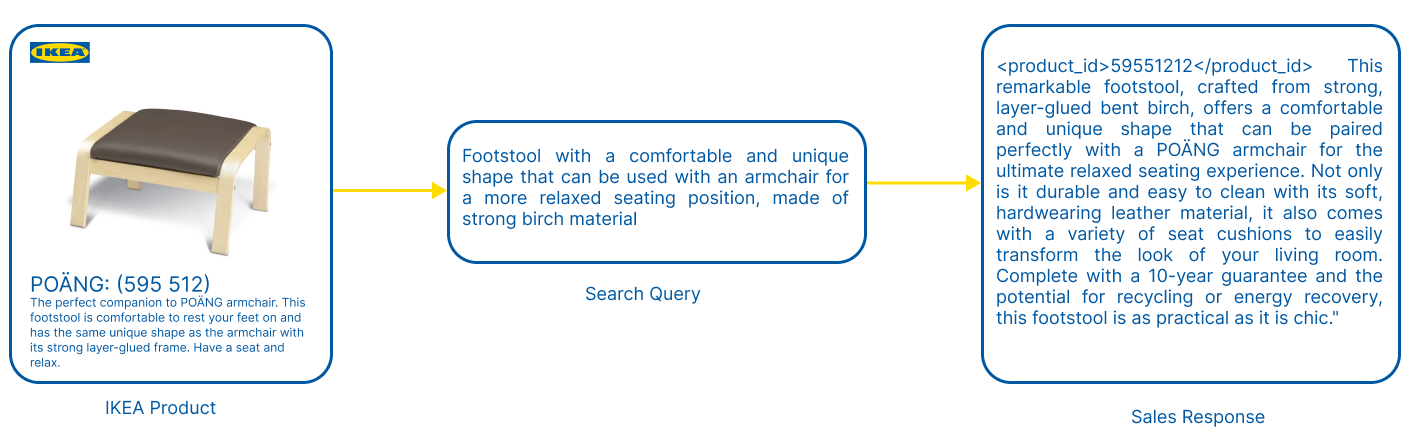}
    \caption{Example of the pipeline from the Product Information to the Sales Response. The LLM fine-tuning is done on the search query and the sales response pair. (\textit{Note: The IKEA logo and the IKEA wordmark are registered trademarks of Inter IKEA Systems B.V.}) }
    \label{fig:pipeline}
\end{figure*}

In this section, we describe the pipeline employed for generating synthetic data pertaining to product knowledge. The sequential steps of this process are also represented visually in Figure \ref{fig:pipeline}.

\subsection{Generating Search Queries}
We initiated our study with a dataset comprising approximately two thousand products sourced from 25 distinct categories within the IKEA inventory. Employing GPT-4, we formulated five distinct search queries characterized by varying degrees of complexity for each individual product, leveraging key attributes such as price, material composition, and descriptive information. A detailed breakdown of the various search query types is elucidated in Table \ref{search_query}, where the descriptions were integral components of the prompts provided to GPT-4.
Ultimately, we had approximately 10,000 search queries in total. 
\begin{table*}[]
    \centering
    \scalebox{0.8}{
    \begin{tabular}{|p{0.5cm}|p{4cm}|p{8cm}|}
    \hline
      \textbf{\#} & \textbf{Type of Search Query} &\textbf{ Description} \\
        \hline
       1.  & Basic & A simple search query that a customer can type on the website \\
       \hline
        2. & Complex & The query can include information such as price, dimension, ease of build, design style such as minimalist but does not have to include all information always. \\
        \hline
        3. & Contextualize & Contextualized search query with context targeted towards \textit{"Audience"} \\
        \hline
        4. & Benefit Query &A search query highlighting benefits based on the provided benefits \\
        \hline
        5. & Contextualize & Contextualized search query with context targeted toward \textit{"New Audience"} \\
        \hline
    \end{tabular}}
    \caption{Different Types of search queries generated for each Product}
    \label{search_query}
\end{table*}

\subsection{Generating the Sales Response}
\begin{figure}
    \centering
    \includegraphics{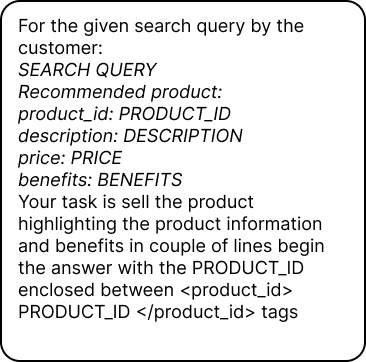}
    \caption{Guided Prompt for generating Sales Response using GPT-4}
    \label{fig:sell-prompt}
\end{figure}

Once we had the multiple search queries for each product, we generated a sales response for the pair of search query and the product knowledge. This generation process was completed using the GPT-4, with the prompt employed depicted in Figure \ref{fig:sell-prompt}.

\subsection{Final Dataset}
In the process of model training, three search queries are randomly allocated for training purposes, and one each is assigned for validation and testing across the entirety of the product inventory. Consequently, the resultant distribution produces 6,099 queries for training, 2,033 for validation, and 2,033 for model evaluation in the designated task.
\section{Methodology}

\subsection{Training}
We perform supervised fine-tuning of LLM using full fine-tuning approaches on the dataset. To ensure that each product is represented by a unique token, namely the product ID, we expand the vocabulary of the LLM so that all product ids are represented as tokens. Furthermore, we also make a comparison without adding product IDs as new tokens. We choose to train Mistral7Bv0.2 \cite{jiang2023mistral} for all our experiments. 
\subsection{Evaluation}
In this section, we detail the evaluation methodology employed to assess the performance of the product recommendation system. Each subsection outlines the specific procedures that were conducted for evaluation. 
\subsubsection{\textbf{Quantitative}} 
\begin{enumerate}
    \item \textbf{Top-1 Match:} We measure the accuracy of the Top-1 recommendations against the ground truth.
    \item \textbf{Top-5 Match:} We measure the accuracy of the Top 5
recommendations against the ground truth. 
\item \textbf{Top-1 Category Match:} We measure the accuracy in terms of the product category of the recommended product against the ground truth.
\item \textbf{Top-5 Category Match:} We measure the accuracy in terms of the product category of the Top 5 recommended product against the ground truth. 
\end{enumerate}

\subsubsection{\textbf{Qualitative: }} In order to evaluate the factual accuracy of the generated sales response of a product Id, we using GPT-4-turbo as a to judge evaluate of few of the following parameters.
\begin{enumerate}
    \item \textbf{Correct-Series-Name}: True/False (if the series name of the product ID in generated response matches the series name of the product ID)
    \item \textbf{Correct-Price}: True/False (if the price of the product ID in generated response matches the price of the product ID)
    \item  \textbf{Relevancy}: True/False (if the product in generated response serves the same purpose as the actual product that belongs the product ID) 
   \item \textbf{Added-New-Information}: True/False (if the generated responses added new information that is not present in the original product description for the product ID)  
   \item \textbf{Correct-Material}: True/False (if the generated response has the same material as the original product description for the product ID )
   \item \textbf{Correct-Color}: True/False (if the generated response has the same color as the original product description for the product ID )
\end{enumerate}
\section{Results}
The Table \ref{tab:results} displays the results of the product recommendation experiments. Interestingly, the model that was trained without extra tokens performed quite well relative to the model that included additional product ID tokens (despite each number being treated as a separate entity in the tokenizer). However, the model that incorporates additional tokens excelled in all evaluation metrics. Furthermore, it was observed that in 3.3\% of recommendations, the model lacking product ID tokens generated the product IDs on its own, while the model with product ID tokens did not exhibit such hallucinations.

\begin{figure*}
    \centering
    \includegraphics{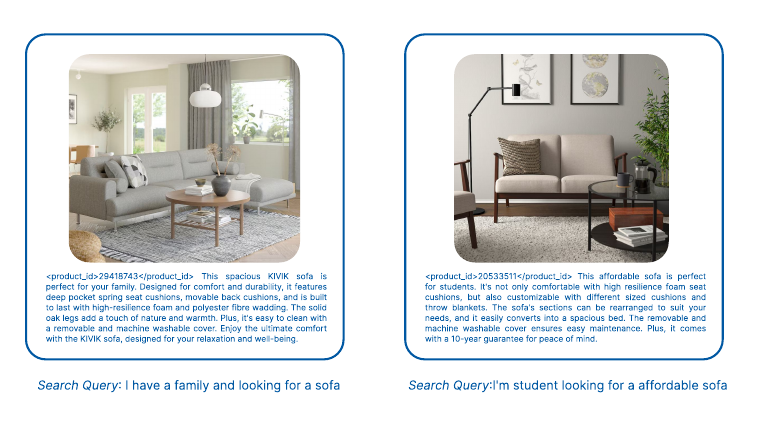}
    \caption{Tailored Sofa Recommendations Based on User Needs: Search Results for Family-Oriented Users (Left) and Budget-Conscious Students (Right).}
    \label{fig:demo}
\end{figure*}

Figure \ref{fig:demo} presents the search results for two distinct user queries: \textit{"I have a family and looking for a sofa"} and \textit{"I'm a student looking for an affordable sofa"}. The results demonstrate the model's ability to understand and cater to the specific needs of the users. For the family-oriented user, the model recommends a spacious 4-person sofa, albeit on the expensive side. In contrast, the model recommends a budget-friendly 2-person sofa for the student user, which also serves as a sofa bed, potentially important for students with limited living space. This highlights the model's ability to generate personalized recommendations based on the unique needs and circumstances of the users.

\begin{table*}[]
    \centering
    \scalebox{0.8}{
    \begin{tabular}{ccccc}
        \textbf{Model} & \textbf{Top-1 Match} & \textbf{Top-5 Match}& \textbf{Top-1 Category Match} & \textbf{Top-5 Category Match} \\
        \hline
         Mistral7B (with new tokens) &  28.7\% & 46.3\% & 91.3\% & 97.1 \%\\
         \hline
         Mistral7B (without new tokens) &  22.7\% & 44.4\% &87.5\% & 96.6\% \\
         \hline
    \end{tabular}}
    \caption{Results of Product Recommendations}
    \label{tab:results}
\end{table*}

\section{Does a LLM learn the product information?}
\begin{figure}
    \scalebox{0.4}{
    \centering
    \includegraphics{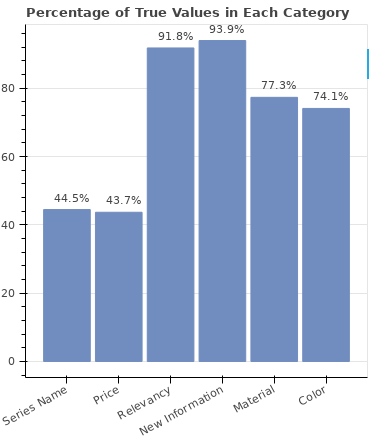}}
    \caption{Results of Qualitative Analysis on Mistral(with Product ID tokens)}
    \label{fig:analysis}
\end{figure}

The evaluation of the product descriptions generated by Mistral(with Product ID tokens) against the original true descriptions was based on six key parameters: Correct-Series Name, Correct Price, Relevancy, New Information Added, Correct Material, and Correct Color. The percentage of true values are shown in the Figure \ref{fig:analysis}.

The model demonstrated a strong understanding of the purpose of the products, with a high Relevancy score of 91.78\%. This indicates that the generated descriptions accurately represented the functionality of the products.

However, the model frequently added new information that was not present in the original product descriptions, with a score of 93.9\%. This suggests that the model has a tendency to generate information that is not factually accurate, which could potentially be misleading.

In terms of factual accuracy, the model's performance was relatively lower. The Correct-Series-Name and Correct-Price scores were 44.4\% and 43.6\% respectively, indicating that the model struggled to accurately generate these specific details.

The model's performance in terms of Correct-Material and Correct-Color was moderate, with scores of 77. 3\% and 74. 0\%, respectively. This suggests that while the model was often able to generate the correct material and color, it also had a significant number of errors in these areas.

\section{Conclusion and Limitations}
In conclusion, this study demonstrates the potential of fine-tuning large language models (LLMs) for product recommendations, with Product IDs incorporated into the vocabulary. The model shows promising results in understanding the purpose of the products and generating contextualized recommendations. However, there are significant areas for improvement. The model struggles with factual accuracy, particularly in relation to the series name and price of the products. Additionally, the model's tendency to add new information that is not present in the original descriptions is a concern, as this information is often not accurate.

Another limitation of this approach is its lack of adaptability to new products. If new products are introduced, the model must be re-trained to adapt to these changes, which could be resource intensive and time-consuming.

To address the issue of hallucination, particularly around the price, we suggest introducing price-specific search queries into the training dataset. Despite these limitations, the model in its current state can still be utilized to generate relevant product IDs, with the sales response generated through retrieval of the product description. The model's performance in generating the correct material and color was moderate, indicating that this is also an area that could be improved.

Future work should focus on enhancing the model's ability to accurately generate factual details and reduce the generation of new, potentially misleading information. Improving the adaptability of the model to new products without the need for retraining would also be a valuable area of research. Additionally, efforts should be made to improve the model's performance in generating correct material and color information.



\bibliographystyle{ACM-Reference-Format}
\bibliography{software.bib}


\end{document}